\documentclass[twocolumn,showpacs,preprintnumbers,amsmath,amssymb, pre]{revtex4}

\usepackage{graphicx}
\usepackage{dcolumn}
\usepackage{bm,xcolor}

\begin{document}


\title{Analytic Continuation of Quantum Monte Carlo Data \\by Stochastic Analytical Inference}

\author{Sebastian Fuchs}
\email{fuchs@theorie.physik.uni-goettingen.de}

\author{Thomas Pruschke}%
\affiliation{%
Institut f\"ur Theoretische Physik, Georg-August-Universit\"at
G\"ottingen, Friedrich-Hund-Platz 1, 37077 G\"ottingen, Germany
}%

\author{Mark Jarrell}
\affiliation{
Louisiana State University, Baton Rouge, Louisiana 70803, USA
}%

\date{\today}

\begin{abstract}
  We present an algorithm for the analytic continuation of
  imaginary-time quantum Monte Carlo data
  which
  is strictly
  based on principles of Bayesian statistical inference. Within this framework
  we are able to obtain an explicit expression for the calculation of a
  weighted average over possible energy spectra, which can be evaluated by
  standard Monte Carlo simulations, yielding as by-product also the distribution
  function as function of the regularization parameter. Our algorithm thus avoids the usual ad-hoc 
  assumptions introduced in similar algortihms to fix the regularization 
  parameter. We apply the algorithm to imaginary-time quantum
  Monte Carlo data and compare the resulting energy spectra with
  those from a standard maximum entropy calculation.
\end{abstract}

\pacs{Insert valid PACS here!}
\maketitle

\section{Introduction}

Quantum Monte Carlo simulations are a powerful computational tool to calculate
properties of interacting quantum manyparticle systems, such as
spin models or strongly correlated electron systems. Of particular
interest in those systems are dynamical correlation functions like
single-particle spectra or susceptibilies respectively dynamical structure
factors. However, QMC presently provides data only on
the imaginary time axis, and the necessary analytic continuation of
these data has proven to be difficult.

The standard tool to solve this problem is the Maximum Entropy Method (MEM)
\cite{maxentrev}. It uses arguments of Bayesian
logic \cite{jaynes,gull} to obtain the most probable energy
spectrum. In order to solve this optimization problem efficiently, the
maximum entropy method approximates all occurring probability
distributions to be of a Gaussian shape. 

In the past efforts were made to provide an alternative to this
approach \cite{sandvik,beach,syljuasen}. It was proposed to perform a
Monte Carlo average over a wide range of spectra instead of selecting
a single spectrum.  So far, the method lacked a rigorous rule to
eliminate a regularization parameter inherent in the
algorithm. Although this approach has been interpreted in terms of
Bayesian inference \cite{syljuasen}, non of the authors have
utililized Bayesian logic to eliminate the regularization parameter.

We show that this stochastic approach can also be understood in terms
of Baysian statistical inference. We derive a strict criterion to
eliminate the free parameter, that is completely based on Bayesian
logic. It uses Monte Carlo techniques to both calculate the average
spectrum and to eliminate the regularization parameter. It treats all
probabilities exactly and hereby avoids the approximations made in the
maximum entropy method. We apply the algorithm to imaginary-frequency
quantum Monte Carlo data and compare the resulting spectra with
results from maximum entropy calculations.

\section{The Problem of Analytic Continuation}

For a finite temperature $T$ quantum Monte Carlo simulations can
provide accurate estimates $\bar G_n$ for either imaginary-time
correlation function $G(\tau)$ at a finite set of $N$ imaginary-time
points $\tau_n$ or, alternatively, for imaginary-frequency correlation
functions $G(i\omega_n)$ at a finite set of $N$ Matsubara frequencies
$\omega_n$. The frequencies are defined as $\omega_n=(2n+1)\pi/\beta$
for fermions and as $\omega_n=2n\pi/\beta$ for bosons with $\beta =
1/k_{\text{B}} T$.

Because of the stochastical nature of Monte Carlo algorithms each of
the $\bar G_n$ possesses a known statistical error. Moreover, the data
for the different time or frequency points are usually highly
correlated. Therefore the input to the analytic continuation procedure
consists of the Monte Carlo estimates $\bar G_i$ and their covariance
matrix
\begin{equation}
  C_{nm} = \overline{G_n G_m} - \bar G_n \bar G_m
\end{equation}
In principle the spectral function $A(\omega) = -\frac{1}{\pi}
\text{Im}\,G(\omega+i0^+)$ can be extracted from these data by
inverting
\begin{equation}\label{eq:rel}
  \bar G_n = \int\! d\omega K_n(\omega) A(\omega)
\end{equation}
with
\begin{equation}
  K_n(\omega) = K(\tau_n,\omega) := -\frac{e^{-\omega\tau}}{1 \pm e^{-\omega\beta}}
\end{equation}
for time dependent data or 
\begin{equation}
  K_n(\omega) = K(i\omega_n, \omega) := \pm \frac{1}{i\omega_n-w}
\end{equation}
for frequency dependent data, where the upper sign holds for fermions
and the lower one for bosons. The spectral function is normalized to
\begin{equation}\label{eq:norm}
  \mathcal{N} = \int\! d\omega A(\omega)
\end{equation}
and is nonnegative for all $\omega$.
However, a direct inversion of Eq.~\ref{eq:rel} is an ill-posed problem and
numerically impossible. 

A least-square fit of $A(\omega)$ to the data $\bar
G_n$ minimizes the \emph{$\chi^2$-estimate}
\begin{equation}
  \chi^2[A] = \frac{1}{N} \sum\limits_{n,m} \left( \bar G_n -
    G(\tau_n) \right)^* \sqrt{C_{nm}^{-1}} \left( \bar G_m -
    G(\tau_m) \right)
\end{equation}
with respect ro $A(\omega)$. This approach leads to a multitude of different
solutions and consequently cannot solve the problem either.

\subsection{The Maximum Entropy Method}

The maximum entropy method can be understood as an attempt to
regularize the least-square fit described above. One defines the
entropy 
\begin{equation}\label{eq:ent}
  S\left[A\right] = -\int\!
  d\omega\,A(\omega) \ln\frac{A(\omega)}{D(\omega)}
\end{equation}
relative to a \emph{default model} $D(\omega)$. Any information, that
is known about the spectrum beforehand, can be encoded in the default
model. If $D(\omega)$ is nonnegative and possesses the same norm
$\mathcal{N}$ as the spectrum $A(\omega)$, the entropy $S$ will be
nonpositiv and maximal for $D(\omega)$. Instead of just minimizing
$\chi^2$ the MEM minimizes the quantity
\begin{equation}\label{eq:qdef}
  Q[A] = \chi^2[A] - \alpha S[A]
\end{equation}
introducing a \emph{regularization parameter} $\alpha$. This
optimization problem can be numerically solved for fixed $\alpha$ to
find the minimizing spectrum $\hat A_{\alpha}(\omega)$. In the limit
of $\alpha \rightarrow \infty$ the spectrum minimizing $Q$ is the
default model $D(\omega)$. For $\alpha \rightarrow 0$ the least-square
fit is regained. Thus the parameter $\alpha$ interpolates between the
fit result and the default model. 

In order to find a criterion to eliminate the parameter different
approaches exist. The simplest rule is to take the spectrum where
$\chi^2\sim 1$. This choice ensures that the differences between model
and data are of the order of the error bars thereby avoiding
overfitting. In order to derive more sophisticated methods the MEM
needs to be reinterpreted by means of Bayesian statistical inference
\cite{jaynes, gull}.

\subsubsection{Bayesian Statistical Inference}

The MEM can be reformulated by defining subjective probabilities for
the quantities involved in the analytic continuation problem. Let
$P[A]$ denote the \emph{prior probability} of the spectrum
$A(\omega)$. $P[A | \bar G]$ denotes the \emph{posterior probability}
of $A$ given the input data $\bar G$ and $P[ \bar G | A]$ the
\emph{likelihood function}. \emph{Bayes's Theorem} \cite{papoulis} relates these probabilities to each
other:
\begin{equation}\label{eq:bayes}
  P[A | \bar G ] = P[ \bar G | A]\ P[A]\ /\ P[\bar G].
\end{equation}
The probability $P[\bar G]$ is called the \emph{evidence} and serves
as normalization for the posterior probability $P[A | \bar G]$:
\begin{equation}\label{eq:evidence}
  P[\bar G] = \int\!\mathcal{D}A\, P[ \bar G | A]\ P[A].
\end{equation}
One indentifies
\begin{equation}\label{eq:hood}
  P[\bar G | A] = \frac{1}{Z_1} \exp(-\chi^2[A])
\end{equation}
and 
\begin{equation}\label{eq:prior}
  P[A] = \frac{1}{Z_2} \exp(\alpha S[A]).
\end{equation}
The quantities
\begin{equation}
  Z_1 = \int\!\mathcal{D}\bar G \, e^{-\chi^2[A]}
\end{equation}
and
\begin{equation}
  Z_2 = \int\!\mathcal{D}A \, e^{\alpha S[A]}
\end{equation}
normalize the respective probabilities.  This way the posterior
probability can be rewritten as
\begin{equation}\label{eq:probspec}
  P[A | \bar G] = \frac{e^{-Q[A]}}{Z_1Z_2P[\bar G]}.
\end{equation}
with 
\begin{equation}
  P[\bar G] = \frac{\int\!\mathcal{D}A\, e^{-Q[A]}}{Z_1Z_2}
\end{equation}
Thus the minimization of $Q$ can be reinterpreted as the maximization
of the posterior probability $P[A | \bar G]\sim e^{-Q}$. The MEM
therefore determines the \emph{most probable} spectrum $\hat
A_{\alpha}$ given the input data $\bar G$.

\subsubsection{Bayesian Inference and the Regularization Parameter $\alpha$}

This alternative formulation of the problem provides the necessary
tools to eliminate the free parameter $\alpha$ \cite{skilling1,
  skilling2}. Eq.~\ref{eq:bayes} can be rewritten including
$\alpha$:
\begin{equation}\label{eq:bayes2}
  P[A, \alpha | \bar G] = P[ \bar G | A, \alpha]\ P[A, \alpha]\ /\ P[\bar G].
\end{equation}
If one applies Bayes's theorem to factorize $P[A,
\alpha]$ and integrates over $A$, the relation
\begin{align}
  P[\alpha| \bar G] =& P[\alpha] \int\!\mathcal{D}A\, P[ \bar G | A,
  \alpha]\ P[A | \alpha]\ /\ P[\bar G] \nonumber\\
  = & \frac{P[\alpha]}{Z_1Z_2P[\bar G]} \int\!\mathcal{D}A\, e^{-Q[A]} \label{eq:probalphamem}
\end{align}
for the posterior probabilty $P[\alpha | \bar G]$ can be found. Analogous to
the argument given above, one identifies $P[\bar G | A, \alpha] \sim \exp(-\chi^2[A])$
and $P[A | \alpha] \sim \exp(\alpha S[A])$. The evidence
\begin{equation}
  P[\bar G] = \int\! d\alpha \frac{P[\alpha]\int\!\mathcal{D}A\, e^{-Q[A]}}{Z_1Z_2} 
\end{equation}
is an $\alpha$-independent normalization constant.  All quantities in
this equation are known except $P[\alpha]$, the prior probability of
$\alpha$. It is either taken to be constant or to be the Jeffreys
prior $1/\alpha$ \cite{markmem, skilling2, bryan}. However, the choice of
$P[\alpha]$ turns out to be of little influence on the resulting
spectra.

By assuming all probabilities involved to be of a Gaussian shape a
numerical treatment of the equations (\ref{eq:probspec}) and
(\ref{eq:probalphamem}) is possible. There are two alternatives:
\begin{enumerate}
\item One calculates $\alpha^*$ as the $\alpha$ that maximizes
  $P[\alpha | \bar G]$ and takes $\hat A_{\alpha^*}$ as the final
  result for the spectral function \cite{skilling1, skilling2}. 
\item One averages over all $\hat A_\alpha$ weighted by the posterior
  probability of $\alpha$, i.e. the average spectrum
  \begin{equation}
    \langle A \rangle = \int\! d\alpha P[\alpha | \bar G] \hat A_{\alpha}
  \end{equation}
  is taken as the final result \cite{bryan}.
\end{enumerate}
It is not a priori clear, which of the two algorithms is favorable.

\subsection{Stochastic Analytical Inference}

Stochastic Analytical Inference is an alternative to the standard MEM which does
not employ the explicit regularization of the fit by the entropy
Eq.~(\ref{eq:ent}). Rather than maximizing $P[A|\bar G]$ an average over
all possible spectra weighted by
\begin{equation}
w \sim \exp(-\chi^2/\alpha)  
\end{equation}
is performed. Beach refined this approach, by introducing the default
model $D(\omega)$ of the MEM into the algorithm \cite{beach}. By
mapping $\omega$ unto $x \in [0,1]$ using
\begin{equation}
  x = \phi(\omega) = \frac{1}{N}\int\limits_{-\infty}^{\omega}\!
  d\omega'\,D(\omega')\;\;, 
\end{equation}
a dimensionless field $n(x)$ can be defined: 
\begin{equation}
  n(x) = \frac{A(\phi^{-1}(x))}{D(\phi^{-1}(x))}\;\;.
\end{equation}
The field $n(x)$ is normalized to 1:
\begin{equation}\label{eq:nnorm}
  1 = \int\limits_0^1\! d x\,n(x)\;\;.
\end{equation}
By calculating the average field
\begin{equation}\label{eq:nav}
  \langle n(x) \rangle_{\alpha} = \frac{1}{Z}\int\!\mathcal{D}'n(x)\, n(x)\,
  e^{-\chi^2[n(x)]/\alpha} 
\end{equation}
with
\begin{equation}
  Z = \int\!\mathcal{D}'n(x)\, e^{-\chi^2[n(x)]/\alpha}\;\;.
\end{equation}
The measure
\begin{equation}\label{eq:measure}
  \mathcal{D}'n(x) = \mathcal{D}n(x) \, \Theta\!\left[n\right] \,
  \delta\!\left(\int\limits_0^1\! d x\,n(x) - 1\right) 
\end{equation}
restricts the integration to fields $n(x)$ that satisfy the norm rule
Eq.~(\ref{eq:nnorm}) and the positivity requirement. In
Eq.~(\ref{eq:measure}) 
$$
\Theta\!\left[n\right]=\left\{\begin{array}{l}\mbox{1, if $\forall x: n(x)\geq 0$}\\
\mbox{$0$ otherwise}\end{array}\right.\;\;.
$$ 
The average spectrum $\langle A \rangle_{\alpha}$ can be regained via
\begin{equation}
  \langle A(\omega) \rangle_{\alpha} = D(\omega) \langle n(\phi(\omega))
  \rangle_{\beta} \;\;.
\end{equation}
If $\chi^2$ is interpreted as a Hamiltonian of a fictitious physical
system, Eq.~(\ref{eq:nav}) possesses the structure of an canonical
ensemble average at a temperature $\alpha$. The laws of
statistical mechanics then state, that the average spectral function
$\langle A \rangle_{\alpha}$ minimizes the free energy
\begin{equation}
  F = \langle \chi^2 \rangle_{\alpha} - \alpha \mathcal{S}\;\;.
\end{equation}
This expression displays a similar structure as Eq.~(\ref{eq:qdef}). Thus
the averaging process implicitely generates an entropy
$\mathcal{S}$. However, this entropy does not have the explicit form
of Eq.~(\ref{eq:ent}). In the limit $\alpha \rightarrow 0$ the
averaging process minimizes $\chi^2$.  Whereas in the limit $\alpha
\rightarrow \infty$ the average in Eq.~(\ref{eq:nav}) is completely uneffected
by $\chi^2$ and will -- constrainded by Eq.~\ref{eq:nnorm} -- result in
$\langle n(x) \rangle = 1$. In this case the resulting
spectrum is the default model. The algorithm therefore exhibits the same
limiting cases as the MEM. Additionaly, Beach has shown that a mean
field treatment of the ficticious physical system described by
$\chi^2$ is formaly equivalent to the MEM \cite{beach}.

The remaining open question, namely how to eliminate the parameter $\alpha$,
 was addressed by all preceding authors differently. 
\begin{enumerate}
\item Sandvik proposes to examine the plot of the average entropy
  against $\alpha$ and identifies the final $\alpha$ by a sharp drop
  in the entropy curve \cite{sandvik}. 
\item Beach examines a double-logarithmic plot of the average
  $\chi^2$ and identifies the the final $\alpha$ by a kink in the
  $\chi^2$-curve \cite{beach}.
\item Sylju\r asen argues to take $\alpha=2N_B/N$, where $N_B$ is the
  number of Monte Carlo bins and $N$ the number of input data points,
  as before \cite{syljuasen}.
\end{enumerate}
All criteria are merely based on heuristic arguments. The simple rule
to take $\chi^2\sim 1$ is also applicable to this method and should be
mentioned here.

\subsubsection{Bayesian Statistical Inference}
In the following
we will use Bayesian inference to derive a new criterion to eliminate the
regularization parameter $\alpha$. In constrast to the MEM the
Stochastic Analytic Continuation does not maximize the posterior
probability $P[A | \bar G]$. Instead, It averages all possible fields
$n$ (omitting the argument $x$ in the progress) weighted by $P[n|\bar
G]$:
\begin{equation}\label{eq:nav2}
  \langle n \rangle = \int\! \mathcal{D}n\, n\, P[n | \bar G]\;\;.
\end{equation}
Bayes's theorem  can be applied to factorize $P[n | \bar G]$
analogous to Eq.~(\ref{eq:bayes}):
\begin{equation}\label{eq:bayesn}
  P[n | \bar G ] = P[ \bar G | n]\ P[n]\ /\ P[\bar G]\;\;.
\end{equation}
The Stochastic Analytic Continuation does not introduce an
explicit entropy term. Following Ref.~\onlinecite{syljuasen} only the positivity
requirement and the norm rule Eq.~(\ref{eq:nnorm}) enter the prior
probability 
\begin{equation}\label{eq:nprior}
  P[n] = \Theta\left(n(x)\right) \,
  \delta\!\left(\int\limits_0^1\! d x\,n(x) - 1\right)\;\;. 
\end{equation}
The likelihood function is identified as
\begin{equation}\label{eq:nlikely}
  P[\bar G | n] = \frac{1}{Z'}e^{-\chi^2/\alpha}\;\;.
\end{equation}
By evaluating a Gaussian integral the normalization $Z'$ can be
readily calculated to be
\begin{equation}\label{eq:zbar}
  Z' = \int\!\mathcal{D}\bar G \, e^{-\chi^2/\alpha} =
  (2\pi\alpha)^{N/2}\sqrt{\det C}\;\;.
\end{equation}
Using 
\begin{equation}
  P[\bar G] = \int\! \mathcal{D}'n \frac{e^{-\chi^2[n]/\alpha}}{Z'} = \frac{Z}{Z'}
\end{equation}
the posterior probability results in
\begin{equation}
  P[ n | \bar G] = \Theta\left(n(x)\right) \,
  \delta\!\left(\int\limits_0^1\! d x\,n(x) - 1\right) \frac{1}{Z}e^{-\chi^2[n]/\alpha},
\end{equation}
as expected from the comparison of the equations (\ref{eq:nav}) and (\ref{eq:nav2}).

\subsubsection{Bayesian Inference and the Regularization Parameter $\alpha$}\label{sssec:regpar}

Bayesian logic can also be utilized to calculate the posterior probability
$P[\alpha | \bar G]$. Substituting $n$ for $A$ in
Eq.~(\ref{eq:probalphamem}) and identifying $P[n | \alpha] = P[n]$ 
with $P[n]$ from Eq.~(\ref{eq:nprior}) and correspondingly
$P[\bar G | n, \alpha] = P[\bar G | n]$ with $P[\bar{G}|n]$ from
Eq.~(\ref{eq:nlikely}), one obtains
\begin{align}
  P[\alpha| \bar G] =& P[\alpha] \int\!\mathcal{D}n\, P[ \bar G | n,
  \alpha]\ P[n | \alpha]\ /\ P[\bar G] \nonumber\\
  = & \frac{P[\alpha]}{Z'P[\bar G]} \int\!\mathcal{D}'n\, e^{-\chi^2[n]/\alpha}\;\;.
\label{eq:probalphasai}
\end{align}
The evidence 
\begin{equation}
  P[\bar G] = \int\! d\alpha\, \frac{P[\alpha] e^{-\chi^2/\alpha}}{Z'(\alpha)}
\end{equation}
is again an $\alpha$-independent normalization constant. The
combination of the equations (\ref{eq:zbar}) and (\ref{eq:probalphasai}) gives
the final expression for the $\alpha$-dependence of the posterior
probability:
\begin{equation}\label{eq:probalpha}
  P[\alpha | \bar G] \sim P[\alpha]\, \alpha^{-N/2} \int\! \mathcal{D}n\, e^{-\chi^2[n]/\alpha}\;\;.
\end{equation}
Analogous to the MEM one has two possibilities to treat the
regularization parameter:
\begin{enumerate}
\item One calculates $\alpha^*$ as the $\alpha$ that maximizes
  $P[\alpha | \bar G]$ and takes $\langle n \rangle_{\alpha^*}$ as the final
  result.
\item One averages over all $\langle n \rangle_\alpha$ weighted by the posterior
  probability of $\alpha$, i.e. the average field
  \begin{equation}
    \langle\langle n \rangle\rangle = \int\! d\alpha P[\alpha | \bar G] \langle n \rangle_{\alpha}
  \end{equation}
  is taken as the final result.
\end{enumerate}

\section{Monte Carlo Evaluation}

\subsection{Configuration and Update Scheme}

In order to calculate the quantities appearing in equations
(\ref{eq:nav}) and (\ref{eq:probalpha}) a numerically treatable
approximation for the field configuration $n(x)$ and the intergration
measure $\mathcal{D}n$ has to be found. Our implementation closely follows
Ref.~\onlinecite{beach}. The field configuration is represented by a
superposition of delta function walkers with residues $r_n$ and
coordinates $x_n$:
\begin{equation}\label{eq:conf}
  n(x) = \sum\limits_{n} r_n \delta(x-x_n).
\end{equation}
The Monte Carlo updates consist of randomly proposed shifts of the
coordinates $x_n$ and random redistributions of the residues
$r_n$. Redistributions that are not only norm conserving but also
conserve higher moments of the configuration \cite{beach}, have proven
to be effective as well.


The average Eq.~(\ref{eq:nav}) is evaluated by a standard Monte Carlo
simulation using Metropolis weights. The regularization parameter
$\alpha$ is treated as the temperature of the system. The simulation
is performed for a wide range of different $\alpha$-values. A
parallel tempering \cite{partemp1, partemp2, partemp3} algorithm is
necessary to ensure convergence for small $\alpha$. In order to
measure the average field configuration a histogram of the delta
function walkers is recorded.

\subsection{Calculation of the probability $P[\alpha | \bar G]$}
A particular problem in the proposed approach is that
a numerical treatment of Eq.~(\ref{eq:probalpha}) involves the calculation
of the quantity
\begin{equation}\label{eq:part}
  Z = \int\!\mathcal{D}'n\, e^{-\chi^2/\alpha}\;\;.
\end{equation}
This is equivalent to calculating a partition function in a canonical
ensemble at temperature $\alpha$. Standard Monte Carlo techniques are
only able to calculate thermal expectation values but not the
partition function itself. We use a \emph{Wang-Landau algorithm}
\cite{wanglandau1, wanglandau2} to generate the density of states
$\rho(E)$ of the system. Once $\rho(E)$ is calculated, the partition
function can be obtained by
\begin{equation}
  Z = \int\!d E\,\rho(E) e^{-E/\alpha}.
\end{equation}
The Wang-Landau algorithm performs a random walk in energy space with
probability $p(E) = 1/\rho(E)$ using the usual metropolis
weights. Since the density of states is unknown at the beginning of
the simulation, one starts with an arbitrary starting value,
e.\,g. $\rho(E)=1$. For each visited energy one updates an energy
histogram and multiplies the density of states by an modication factor
$f>1$.  When the histogram is reasonably flat, one resets the
histogram and restarts the simulation with a new modification factor
$f' = \sqrt{f}$. The starting value of $f$ is usually taken to be
Euler's constant and the procedure is repeated until $f$ is very close to
1 (16 times in our implementation). The resulting $\rho(E)$ is the
density of states of the system up to an unknown normalization
factor. In order to speed up the convergence of the algorithm, it is
advisable to divide the energy range of interest into several slightly
overlapping smaller intervalls.

\section{Simulation Results}

\subsection{The model}
We apply the algorithm to imaginary-time data from quantum Monte Carlo
simulations. As test case we consider the two-dimensional single-band Hubbard model
\begin{equation}
  H = -t \sum\limits_{\langle i,j\rangle \sigma} c^{\dagger}_{i\sigma}
  c_{j\sigma} + U \sum\limits_{i} n_{i\uparrow} n_{i\downarrow}\;\;.
\end{equation}
Here $i$ and $j$ are lattice site indices, the operators
$c^{\dagger}_{i\sigma}$ ($c_{i\sigma}$) create (destroy) an electron
with spin $\sigma \in \{\uparrow, \downarrow\}$ at site $i$,
$n_{i\sigma} = c^{\dagger}_{i\sigma}c_{i\sigma}$ is their corresponding
number density, $t$ is the hopping parameter between neighbouring sites
(denoted by $\langle i, j \rangle$) and $U$ implements the local Coulomb
repulsion. The full lattice model was approximated by a two by two
cluster embedded in a mean field using the Dynamical Cluster
Approximation \cite{hettler1, hettler2, clusterreview}. Using a
weak-coupling expansion in continuous imaginary time \cite{rubtsov1,
  rubtsov2} the single-particle Green function
\begin{equation}
  G(i\omega_n) = - \int\limits_0^{\beta} \!d\tau\, e^{i\omega_n\tau} \langle
  \mathcal{T} c_i(\tau) c^{\dagger}_i \rangle  
\end{equation}
was calculated for a certain number of Matsubara frequencies
$\omega_n=(2n+1)\pi/\beta$. Here $\mathcal{T}$ is the imaginary-time
ordering operator, $\langle\cdot\rangle$ denotes a thermal expectation
value and $c_i(\tau) = e^{-H\tau} c_i e^{H\tau}$. The model was
simulated for $U = W$, where $W=8\,t$ denotes the bandwidth, and a
fixed filling $\langle n_i \rangle=0.9$ for several temperatures
$T$. Within the weak coupling expansion it is possible to calculate
the Green function directly in frequency space \cite{rubtsov2}, so
that no Fourier transformation or discretization of the imaginary time
axis is necessary. In all simulations the number of measured matsubara
frequencies was restricted to $n_{\text{max}}=2U\beta$, which has
proven to be sufficient for all calculation. A further increase of the
number of frequencies had no influence on the analytic continuation
results.

\subsection{Monte Carlo results}

\begin{figure}
  \begin{center}
    \includegraphics{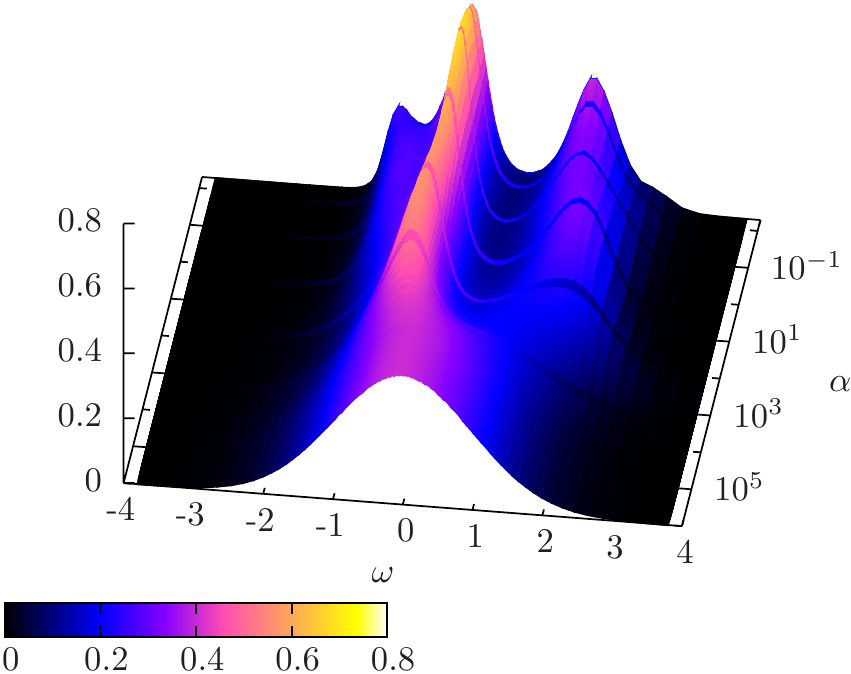}
    \caption{Simulated spectra for a range of regularization
      parameters $\alpha$ ($\beta=14\,W^{-1}$). For large $\alpha$ the Gaussian shape of the
      default model is visible. For decreasing $\alpha$
      several features begin to appear.}
    \label{fig:spectra}
  \end{center}
\end{figure}
Fig~\ref{fig:spectra} shows the $\alpha$-dependence of the single
particle spectra calculated by the parallel tempering Monte Carlo
simulation ($\beta=14\,W^{-1}$). A Gaussian default model 
\begin{equation}\label{eq:gauss}
  D(\omega) = \frac{1}{\sqrt{2\pi\sigma}} e^{-\omega^2/2\sigma}
\end{equation}
with $\sigma = 1$ was used.  The shape of the default model is clearly
visible for large $\alpha$. One can see how several different peaks
and other structures appear for for decreasing $\alpha$. Since the
$\alpha$-dependence is so strong one definitely needs a criterion to
eliminate the regularization parameter.

\begin{figure}
  \begin{center}
    \includegraphics{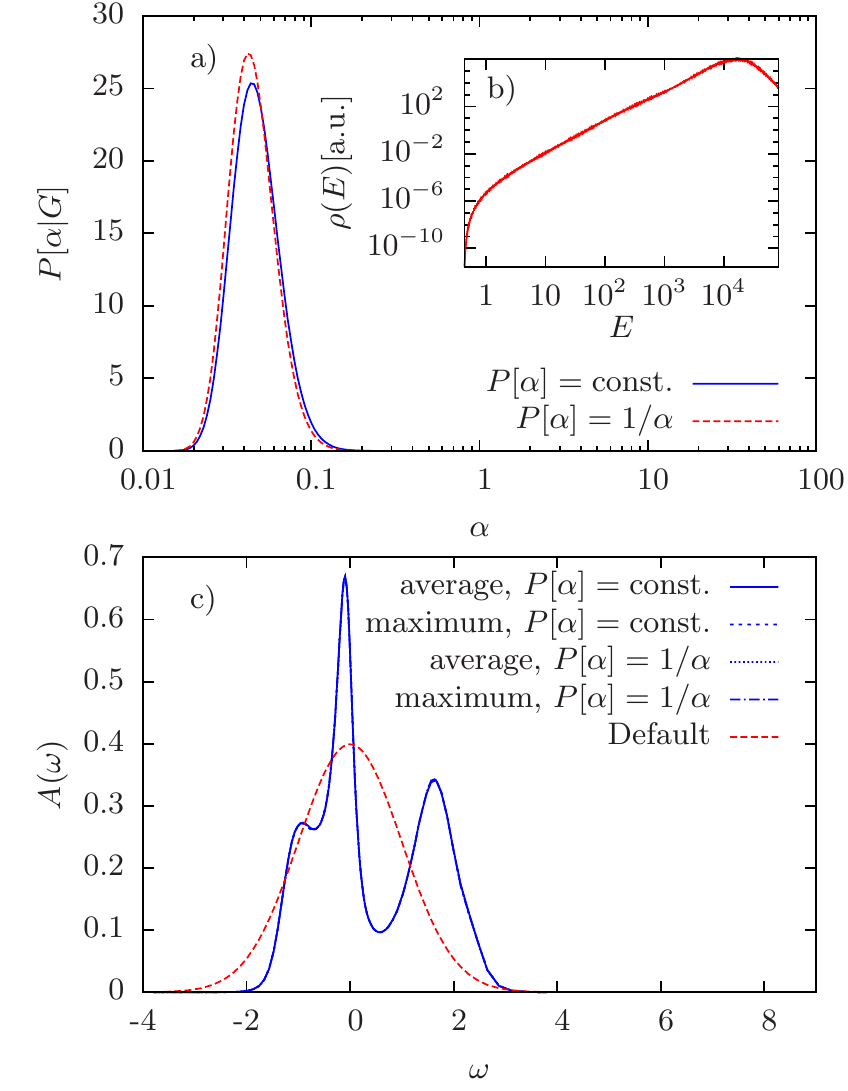}
    \caption{The probability distributions $P[\alpha | \bar G]$ (a)
      based on a Wang-Landau simulation of the density of states
      (b). The different choices for $P[\alpha]$ only have a weak
      influence on the position of the peak. The resulting spectra (c)
      are calculated by either averaging all spectra over $P[\alpha |
      \bar G]$ or by taking the spectrum that maximizes them. The four
      different spectra are practically identical.}
    \label{fig:wanglandau} 
  \end{center}
\end{figure}
The density of states calculated by the Wang-Landau simulation and the
probability distribution $P[\alpha | \bar G]$ following
Eq~(\ref{eq:probalpha}) is shown in Fig.~\ref{fig:wanglandau}. $P[\alpha
| \bar G]$ is plotted for the two most common choices for $P[\alpha]$,
i.\,e. $P[\alpha]=\mathrm{const.}$ and $P[\alpha]=1/\alpha$. The
density of states varies over at least 15 orders of magnitude (note
the logarithmic scales). The probability distributions $P[\alpha |
\bar G]$ exhibit a well defined peak at $\hat\alpha \sim 0.03$. Note that the
two different choices for $P[\alpha]$ have only weak influence on the
position of the peak. The two different probability distributions are
used to calculate the final single particle spectrum. Following the
discussion in section~\ref{sssec:regpar}, Fig.~\ref{fig:wanglandau}
shows the average of all spectra of Fig~\ref{fig:spectra}
weighted by $P[\alpha | \bar G]$ and the spectrum whose $\alpha$
maximizes $P[\alpha | \bar G]$. The resulting spectra are nearly
indistinguishable and show that neither the ambiguity in the treatment
of the probability distribution nor the choice of $P[\alpha]$ have a
significant influence on the resulting spectrum.

Let us compare our results from the stochastic analytical inference with
those obtained with other methods to fix $\alpha$.
\begin{figure}
  \begin{center}
    \includegraphics{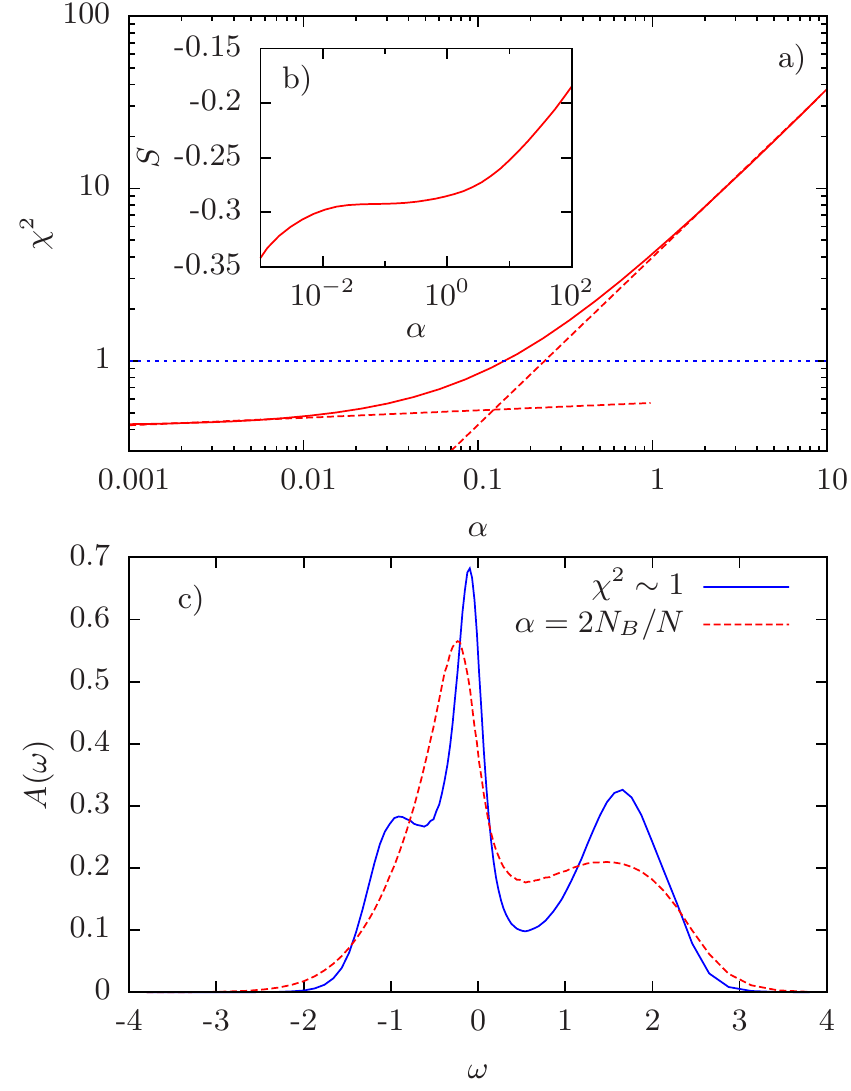}
    \caption{ The double logarithmic plot of $\chi^2$ (a) shows a kink
      at $\alpha\sim 0.1$ which is also the region where $\chi^2\sim
      1$. The resulting spectrum (c) is similar to the one in 
      \ref{fig:wanglandau}b.  The entropy (b) shows no significant
      features and gives at least for this data set no indication how
      to determine $\alpha$. The spectrum for $\alpha = 2N_B/N \sim
      140$ (c) is much more regularized than the spectra determined by
      all other methods.}
    \label{fig:alt}
  \end{center}
\end{figure}
Fig.~\ref{fig:alt} shows that the point where $\chi^2\sim 1$
corresponds to $\alpha\sim 0.1$ and that the $\chi^2$-estimate exhibits
a kink in the same $\alpha$-region. The chosen $\alpha=0.1$ is larger
than $\hat\alpha \sim 0.03$. That indicates that the spectra
determined with this criterion are stronger regularized than the
spectra calculated by the probability distributions in
Fig.~\ref{fig:wanglandau}. However, at least for the QMC data under
consideration, the difference between the two spectra are only
small. The spectrum for $\alpha = \frac{2N_B}{N}$ is also shown in
Fig~\ref{fig:alt}. For the present data set this corresponds to
$\alpha \sim 140$ which is far more regularized than the spectra
determined by all other methods. The entropy (Fig.~\ref{fig:alt}b)
shows no significant features and gives no indication how to choose
the $\alpha$-parameter. A sharp drop in the entropy curve is not
visible in the simulated area.


Finally, we compare the SAI with the standard MEM approach.
\begin{figure}
  \begin{center}
    \includegraphics{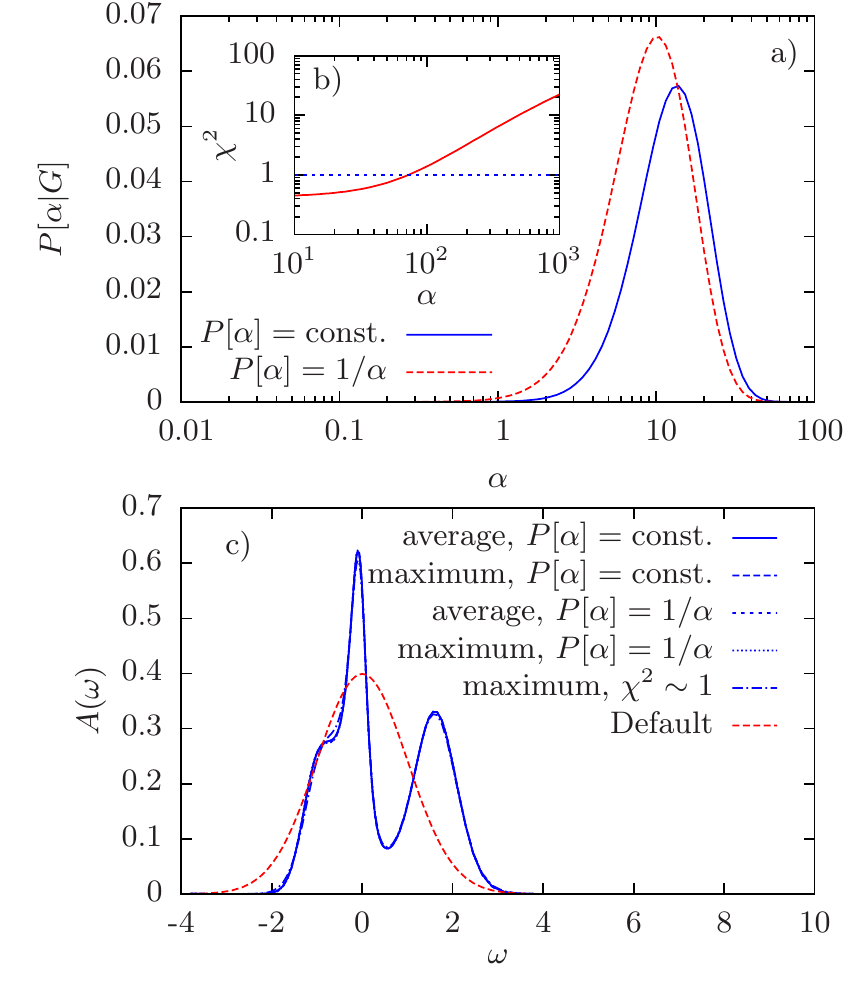}\\
    \caption{Results of a Maximum Entropy calculation for the same QMC
      data as in Fig.~\ref{fig:wanglandau} and
      Fig.~\ref{fig:alt}. The probability distribution $P[\alpha |
      \bar G]$ (a) shows a noticeable dependence on $P[\alpha]$, but
      analogous to SAI, the resulting spectra (c) are identical. The
      $\alpha$ where $\chi^2\sim 1$ (b) is larger ($\alpha\sim 100$)
      than the one for which $P[\alpha | \bar G]$ is maximal ($\alpha
      \sim 10$). Accordingly the spectrum chosen by the $\chi^2\sim 1$
      rule is more regularized than the one calculated by Bayesian
      inference.}
    \label{fig:maxent}
  \end{center}
\end{figure}
Fig.~\ref{fig:maxent} shows results of a Maximum Entropy calculation
using Bryan's algorithm \cite{bryan} for the same QMC data as in the
previous section. The qualitative behaviour is similar to the SAI
simulation: The probability distribution $P[\alpha | \bar G]$ shows a
noticeable dependence on the prior probability $P[\alpha]$. However,
the resulting spectra depend neither on $P[\alpha]$ nor on wether one
averages over $P[\alpha | \bar G]$ or wether one takes the
maximum. The $\chi^2\sim 1$ rule determines an $\alpha$ which is again
larger than the one calculated by Bayesian inference. Accordingly the
spectrum calculated by this criterion is more regularized, although
here the difference is relatively small. Interestingly, in MEM the
interesting values for $\alpha$
are about one or two orders of magnitude larger compared to
those appearing in the SAI simulations. 
There seems to be no direct correspondence between
the $\alpha$-values of the two methods.

\begin{figure*}
  \begin{center}
    \includegraphics{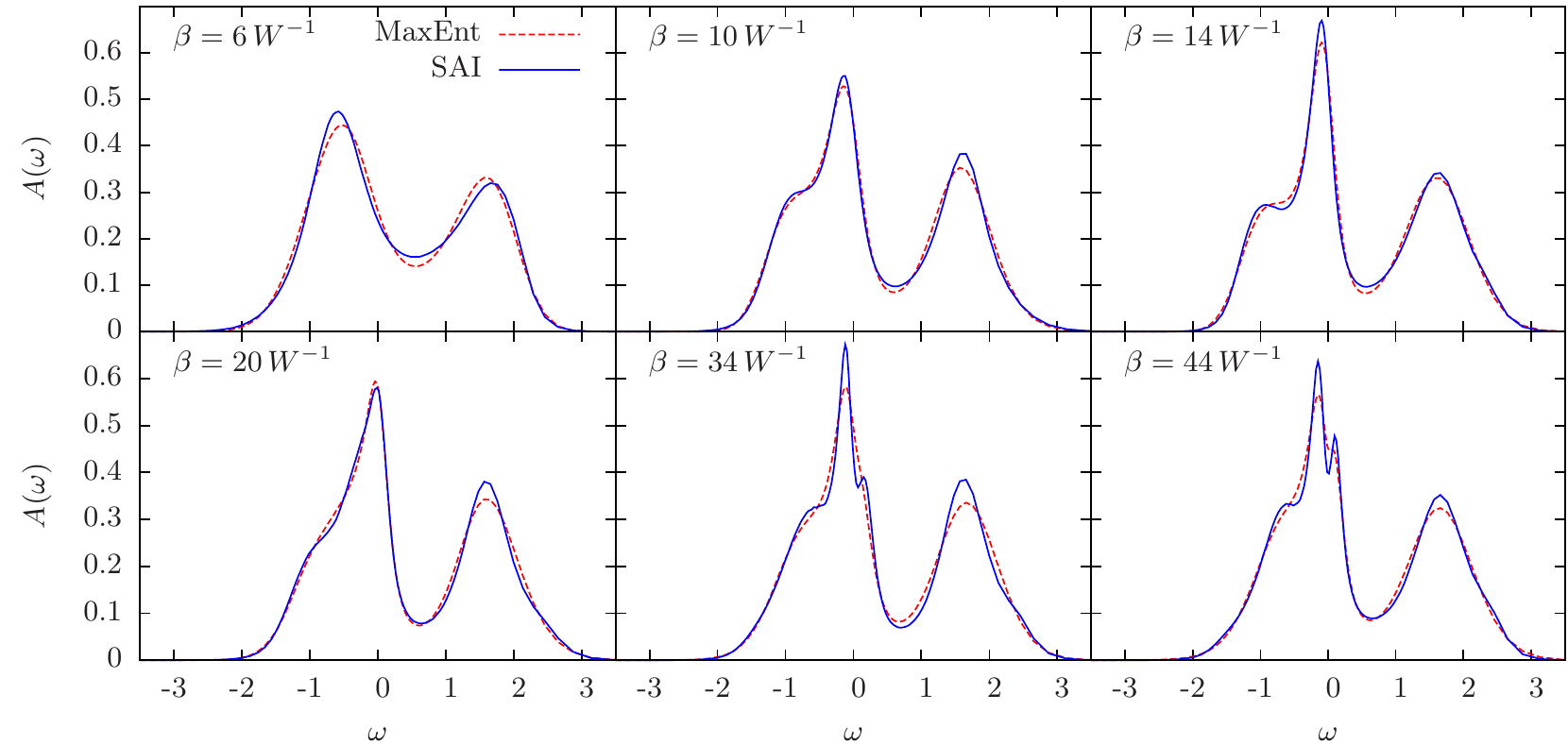}\\
    \caption{Spectra simulated by Stochastic Analytic Inference
      compared to Maximum Entropy Calculations. All calculations are
      based on a Gaussian default model}
    \label{fig:gaussian}
  \end{center}
\end{figure*}
An extended comparison of SAI spectra with results of maximum entropy
calculations for several temperatures is collected in
Fig.~\ref{fig:gaussian}. All calculations are based on the Gaussian
default model Eq~(\ref{eq:gauss}). As already noted before, MEM tends to stronger
regularize the spectra and consequently do the SAI spectra exhibit noticeably
sharper features for all temperatures shown. Especially the
pseudo-gap, that opens at $beta=34\,W^{-1}$, is captured nicely by SAI while
the MEM cannot resolve it yet at that temperature.

One last, but nevertheless important, questions concerns the dependence of
\begin{figure}
  \begin{center}
    \includegraphics{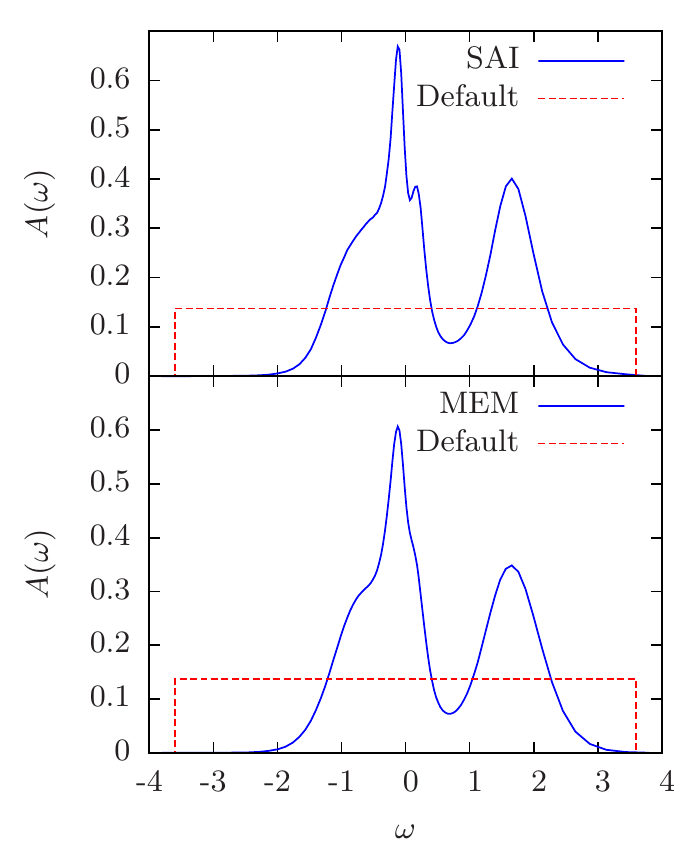}
    \caption{A Stochastic Analytic Inference result for $\beta=34\,W^{-1}$
      based on a flat default model. The spectrum is very
      similar to the one shown in Fig.~\ref{fig:wanglandau}c. We conclude,
      that for the QMC data under consideration the calculated spectra
      are quite independent of the default model. }
    \label{fig:flat} 
  \end{center}
\end{figure}
the spectra on the default model. To this end we show in 
Fig.~\ref{fig:flat} again SAI and the MEM results for the spectrum at
$\beta=34\,W^{-1}$, this time however based on a different default model,
namely a rectangular default model of width $3.6$. The 
resulting spectra are very similar to
the one obtained for the Gaussian default model presented
in Fig.~\ref{fig:gaussian}. Thus, even 
at low temperatures the resulting spectra are quite independent of
the default model. More precisely, we could not detect a significant default model
dependence at any temperature.

\section{Conclusion}

We have demonstrated that the stochastic analytic continuation method introduced by
Sandvik and Beach can be interpreteted in terms of Bayesian
probability theory. We developed an algorithm that uses Monte Carlo
techniques to both calculate the average spectrum and to eliminate the
regularization parameter. It treats all probabilities exactly and
hereby avoids the approximations made in the maximum entropy method.

Comparisons to different approaches to fix the regulariazation parameter $\alpha$
and standard MEM 
show that the SAI results in robust spectral functions which are less regularized
and consequently show more pronounced features, in particular with decreasing
temperature in the model calculations. As known from standard MEM, no significant
dependence on the default model could be observed. 

On apparent drawback of the method is the necessity to perform simulations for
a broad range of values for $\alpha$. Although this can be performed with parallel
tempering techniques, the required computer resources are orders of magnitude
larger than for standard MEM approaches. As the resulting spectra tend to
be less regularized one has to ponder the gain in details in the structures
against the dramatic increase in computer time.

\appendix

\section*{Acknowledgements}

Our implementation of all Monte Carlo algorithms is based on the
libraries of the ALPS project \cite{alps}. ALPS (Applications and
Libraries for Physics Simulations) is an open source effort providing
libraries and simulation codes for strongly correlated quantum
mechanical systems.

We acknowledge financial support by the Deutsche
Forschungsgemeinschaft through SFB 602 and by the German Academic
Exchange Service (DAAD).

\begin{thebibliography}{22}
\expandafter\ifx\csname natexlab\endcsname\relax\def\natexlab#1{#1}\fi
\expandafter\ifx\csname bibnamefont\endcsname\relax
  \def\bibnamefont#1{#1}\fi
\expandafter\ifx\csname bibfnamefont\endcsname\relax
  \def\bibfnamefont#1{#1}\fi
\expandafter\ifx\csname citenamefont\endcsname\relax
  \def\citenamefont#1{#1}\fi
\expandafter\ifx\csname url\endcsname\relax
  \def\url#1{\texttt{#1}}\fi
\expandafter\ifx\csname urlprefix\endcsname\relax\def\urlprefix{URL }\fi
\providecommand{\bibinfo}[2]{#2}
\providecommand{\eprint}[2][]{\url{#2}}

\bibitem[{\citenamefont{Jarrell and Gubernatis}(1996)}]{maxentrev}
\bibinfo{author}{\bibfnamefont{M.}~\bibnamefont{Jarrell}} \bibnamefont{and}
  \bibinfo{author}{\bibfnamefont{J.}~\bibnamefont{Gubernatis}},
  \bibinfo{journal}{Phys.\ Rep.} \textbf{\bibinfo{volume}{269}},
  \bibinfo{pages}{133} (\bibinfo{year}{1996}).

\bibitem[{\citenamefont{Jaynes}(1986)}]{jaynes}
\bibinfo{author}{\bibfnamefont{E.~T.} \bibnamefont{Jaynes}}, in
  \emph{\bibinfo{booktitle}{Maximum-Entropy and Bayesian Methods in Applied
  Statistics}}, edited by \bibinfo{editor}{\bibfnamefont{J.~H.}
  \bibnamefont{Justice}} (\bibinfo{publisher}{Cambridge University Press},
  \bibinfo{address}{Cambridge}, \bibinfo{year}{1986}), p.~\bibinfo{pages}{1}.

\bibitem[{\citenamefont{Gull}(1988)}]{gull}
\bibinfo{author}{\bibfnamefont{S.~F.} \bibnamefont{Gull}}, in
  \emph{\bibinfo{booktitle}{Maximum-Entropy and Bayesian Methods in Science and
  Engineering}}, edited by \bibinfo{editor}{\bibfnamefont{G.~J.}
  \bibnamefont{Erickson}} \bibnamefont{and}
  \bibinfo{editor}{\bibfnamefont{C.~R.} \bibnamefont{Smith}}
  (\bibinfo{publisher}{Kluwer Academic Press}, \bibinfo{address}{Dordrecht},
  \bibinfo{year}{1988}), vol.~\bibinfo{volume}{1}, p.~\bibinfo{pages}{53}.

\bibitem[{\citenamefont{Sandvik}(1998)}]{sandvik}
\bibinfo{author}{\bibfnamefont{A.~W.} \bibnamefont{Sandvik}},
  \bibinfo{journal}{Phys. Rev. B} \textbf{\bibinfo{volume}{57}},
  \bibinfo{pages}{10287} (\bibinfo{year}{1998}).

\bibitem[{\citenamefont{Beach}(2004)}]{beach}
\bibinfo{author}{\bibfnamefont{K.}~\bibnamefont{Beach}} (\bibinfo{year}{2004}),
  \eprint{cond-mat/0403055}.

\bibitem[{\citenamefont{Sylju{\r a}sen}(2008)}]{syljuasen}
\bibinfo{author}{\bibfnamefont{O.~F.} \bibnamefont{Sylju{\r a}sen}},
  \bibinfo{journal}{Phys. Rev. B} \textbf{\bibinfo{volume}{78}},
  \bibinfo{pages}{174429} (\bibinfo{year}{2008}).

\bibitem[{\citenamefont{Papoulis}(1990)}]{papoulis}
\bibinfo{author}{\bibfnamefont{A.}~\bibnamefont{Papoulis}},
  \emph{\bibinfo{title}{Probability and Statistics}}
  (\bibinfo{publisher}{Prentice Hall}, \bibinfo{address}{New York},
  \bibinfo{year}{1990}), p. \bibinfo{pages}{422}.

\bibitem[{\citenamefont{Skilling}(1989{\natexlab{a}})}]{skilling1}
\bibinfo{author}{\bibfnamefont{J.}~\bibnamefont{Skilling}}, in
  \emph{\bibinfo{booktitle}{Maximum Entropy and Bayesian Methods}}, edited by
  \bibinfo{editor}{\bibfnamefont{J.}~\bibnamefont{Skilling}}
  (\bibinfo{publisher}{Kluwer Academic Press}, \bibinfo{address}{Dordrecht},
  \bibinfo{year}{1989}{\natexlab{a}}), p.~\bibinfo{pages}{45}.

\bibitem[{\citenamefont{Skilling}(1989{\natexlab{b}})}]{skilling2}
\bibinfo{author}{\bibfnamefont{J.}~\bibnamefont{Skilling}}, in
  \emph{\bibinfo{booktitle}{Maximum Entropy and Bayesian Methods}}, edited by
  \bibinfo{editor}{\bibfnamefont{J.}~\bibnamefont{Skilling}}
  (\bibinfo{publisher}{Kluwer Academic Press}, \bibinfo{address}{Dordrecht},
  \bibinfo{year}{1989}{\natexlab{b}}), p.~\bibinfo{pages}{53}.

\bibitem[{\citenamefont{Gubernatis et~al.}(1991)\citenamefont{Gubernatis,
  Jarrell, Silver, and Sivia}}]{markmem}
\bibinfo{author}{\bibfnamefont{J.~E.} \bibnamefont{Gubernatis}},
  \bibinfo{author}{\bibfnamefont{M.}~\bibnamefont{Jarrell}},
  \bibinfo{author}{\bibfnamefont{R.~N.} \bibnamefont{Silver}},
  \bibnamefont{and} \bibinfo{author}{\bibfnamefont{D.~S.} \bibnamefont{Sivia}},
  \bibinfo{journal}{Phys. Rev. B} \textbf{\bibinfo{volume}{44}},
  \bibinfo{pages}{6011} (\bibinfo{year}{1991}).

\bibitem[{\citenamefont{Bryan}(1990)}]{bryan}
\bibinfo{author}{\bibfnamefont{R.~K.} \bibnamefont{Bryan}},
  \bibinfo{journal}{Eur. Biophys. J.} \textbf{\bibinfo{volume}{18}},
  \bibinfo{pages}{165} (\bibinfo{year}{1990}).

\bibitem[{\citenamefont{Swendsen and Wang}(1986)}]{partemp1}
\bibinfo{author}{\bibfnamefont{R.~H.} \bibnamefont{Swendsen}} \bibnamefont{and}
  \bibinfo{author}{\bibfnamefont{J.}~\bibnamefont{Wang}},
  \bibinfo{journal}{Phys. Rev. Lett.} \textbf{\bibinfo{volume}{57}},
  \bibinfo{pages}{2607} (\bibinfo{year}{1986}).

\bibitem[{\citenamefont{Lyubartsev et~al.}(1992)\citenamefont{Lyubartsev,
  Martsinovski, Shevkunov, and Vorontsov-Vel'yaminov}}]{partemp2}
\bibinfo{author}{\bibfnamefont{A.~P.} \bibnamefont{Lyubartsev}},
  \bibinfo{author}{\bibfnamefont{A.~A.} \bibnamefont{Martsinovski}},
  \bibinfo{author}{\bibfnamefont{S.~V.} \bibnamefont{Shevkunov}},
  \bibnamefont{and} \bibinfo{author}{\bibfnamefont{P.~N.}
  \bibnamefont{Vorontsov-Vel'yaminov}}, \bibinfo{journal}{J. Chem. Phys.}
  \textbf{\bibinfo{volume}{96}}, \bibinfo{pages}{1776} (\bibinfo{year}{1992}).

\bibitem[{\citenamefont{Marinari and Parisi}(1992)}]{partemp3}
\bibinfo{author}{\bibfnamefont{E.}~\bibnamefont{Marinari}} \bibnamefont{and}
  \bibinfo{author}{\bibfnamefont{G.}~\bibnamefont{Parisi}},
  \bibinfo{journal}{Europhys. Lett.} \textbf{\bibinfo{volume}{19}},
  \bibinfo{pages}{451} (\bibinfo{year}{1992}).

\bibitem[{\citenamefont{Wang and Landau}(2001{\natexlab{a}})}]{wanglandau1}
\bibinfo{author}{\bibfnamefont{F.}~\bibnamefont{Wang}} \bibnamefont{and}
  \bibinfo{author}{\bibfnamefont{D.~P.} \bibnamefont{Landau}},
  \bibinfo{journal}{Phys. Rev. Lett.} \textbf{\bibinfo{volume}{86}},
  \bibinfo{pages}{2050} (\bibinfo{year}{2001}{\natexlab{a}}).

\bibitem[{\citenamefont{Wang and Landau}(2001{\natexlab{b}})}]{wanglandau2}
\bibinfo{author}{\bibfnamefont{F.}~\bibnamefont{Wang}} \bibnamefont{and}
  \bibinfo{author}{\bibfnamefont{D.~P.} \bibnamefont{Landau}},
  \bibinfo{journal}{Phys. Rev. E} \textbf{\bibinfo{volume}{64}},
  \bibinfo{pages}{056101} (\bibinfo{year}{2001}{\natexlab{b}}).

\bibitem[{\citenamefont{Hettler et~al.}(1998)\citenamefont{Hettler,
  Tahvildar-Zadeh, Jarrell, Pruschke, and Krishnamurthy}}]{hettler1}
\bibinfo{author}{\bibfnamefont{M.~H.} \bibnamefont{Hettler}},
  \bibinfo{author}{\bibfnamefont{A.~N.} \bibnamefont{Tahvildar-Zadeh}},
  \bibinfo{author}{\bibfnamefont{M.}~\bibnamefont{Jarrell}},
  \bibinfo{author}{\bibfnamefont{T.}~\bibnamefont{Pruschke}}, \bibnamefont{and}
  \bibinfo{author}{\bibfnamefont{H.~R.} \bibnamefont{Krishnamurthy}},
  \bibinfo{journal}{Phys. Rev. B} \textbf{\bibinfo{volume}{58}},
  \bibinfo{pages}{R7475} (\bibinfo{year}{1998}).

\bibitem[{\citenamefont{Hettler et~al.}(2000)\citenamefont{Hettler, Mukherjee,
  Jarrell, and Krishnamurthy}}]{hettler2}
\bibinfo{author}{\bibfnamefont{M.~H.} \bibnamefont{Hettler}},
  \bibinfo{author}{\bibfnamefont{M.}~\bibnamefont{Mukherjee}},
  \bibinfo{author}{\bibfnamefont{M.}~\bibnamefont{Jarrell}}, \bibnamefont{and}
  \bibinfo{author}{\bibfnamefont{H.~R.} \bibnamefont{Krishnamurthy}},
  \bibinfo{journal}{Phys. Rev. B} \textbf{\bibinfo{volume}{61}},
  \bibinfo{pages}{12739} (\bibinfo{year}{2000}).

\bibitem[{\citenamefont{Maier et~al.}(2005)\citenamefont{Maier, Jarrell,
  Pruschke, and Hettler}}]{clusterreview}
\bibinfo{author}{\bibfnamefont{T.}~\bibnamefont{Maier}},
  \bibinfo{author}{\bibfnamefont{M.}~\bibnamefont{Jarrell}},
  \bibinfo{author}{\bibfnamefont{T.}~\bibnamefont{Pruschke}}, \bibnamefont{and}
  \bibinfo{author}{\bibfnamefont{M.~H.} \bibnamefont{Hettler}},
  \bibinfo{journal}{Rev. Mod. Phys.} \textbf{\bibinfo{volume}{77}},
  \bibinfo{pages}{1027} (\bibinfo{year}{2005}).

\bibitem[{\citenamefont{Rubtsov and Lichtenstein}(2004)}]{rubtsov1}
\bibinfo{author}{\bibfnamefont{A.~N.} \bibnamefont{Rubtsov}} \bibnamefont{and}
  \bibinfo{author}{\bibfnamefont{A.~I.} \bibnamefont{Lichtenstein}},
  \bibinfo{journal}{JETP Lett.} \textbf{\bibinfo{volume}{80}},
  \bibinfo{pages}{61} (\bibinfo{year}{2004}).

\bibitem[{\citenamefont{Rubtsov et~al.}(2005)\citenamefont{Rubtsov, Savkin, and
  Lichtenstein}}]{rubtsov2}
\bibinfo{author}{\bibfnamefont{A.~N.} \bibnamefont{Rubtsov}},
  \bibinfo{author}{\bibfnamefont{V.~V.} \bibnamefont{Savkin}},
  \bibnamefont{and} \bibinfo{author}{\bibfnamefont{A.~I.}
  \bibnamefont{Lichtenstein}}, \bibinfo{journal}{Phys. Rev. B}
  \textbf{\bibinfo{volume}{72}}, \bibinfo{pages}{035122}
  (\bibinfo{year}{2005}).

\bibitem[{\citenamefont{Albuquerque et~al.}(2007)\citenamefont{Albuquerque,
  Alet, Corboz, Dayal, Feiguin, Fuchs, Gamper, Gull, G{\"u}rtler, Honecker
  et~al.}}]{alps}
\bibinfo{author}{\bibfnamefont{A.~F.} \bibnamefont{Albuquerque}},
  \bibinfo{author}{\bibfnamefont{F.}~\bibnamefont{Alet}},
  \bibinfo{author}{\bibfnamefont{P.}~\bibnamefont{Corboz}},
  \bibinfo{author}{\bibfnamefont{P.}~\bibnamefont{Dayal}},
  \bibinfo{author}{\bibfnamefont{A.}~\bibnamefont{Feiguin}},
  \bibinfo{author}{\bibfnamefont{S.}~\bibnamefont{Fuchs}},
  \bibinfo{author}{\bibfnamefont{L.}~\bibnamefont{Gamper}},
  \bibinfo{author}{\bibfnamefont{E.}~\bibnamefont{Gull}},
  \bibinfo{author}{\bibfnamefont{S.}~\bibnamefont{G{\"u}rtler}},
  \bibinfo{author}{\bibfnamefont{A.}~\bibnamefont{Honecker}},
  \bibnamefont{et~al.}, \bibinfo{journal}{Journal of Magnetism and Magnetic
  Materials} \textbf{\bibinfo{volume}{310}}, \bibinfo{pages}{1187}
  (\bibinfo{year}{2007}), \eprint{http://alps.comp-phys.org}.

\end{thebibliography}

\end{document}